**Full-body deep learning-based automated contouring of contrast-enhanced murine organs for small animal irradiator CBCT**

**Authors:** Ethan Cramer[1], Sophie Dobiasch[2,3], Xinmin Liu[1], Stephanie E. Combs[2,3], Rodney D. Wiersma[1]

**Affiliations:**
1. Department of Radiation Oncology, University of Pennsylvania, Philadelphia, PA, USA
2. Department of Radiation Oncology, Technical University of Munich, Munich, Germany
3. Institute of Radiation Medicine (IRM), Helmholtz Zentrum München, Neuherberg, Germany

**Running Title:**
- DL contrast-enhanced murine contouring for CBCT

**Abstract**

*Purpose.* To alleviate the manual contouring burden, deep learning (DL) based automated contouring has been explored. However, due to the poor contrast resolution of preclinical irradiator CBCT, these methods have been limited to high contrast—minimally anatomically complex—structures such as the heart and lungs. Thus, low contrast abdominal CBCT DL-based segmentation has yet to be addressed. In this work we explore a DL-based model in conjunction with iodine-based contrast agent approach to allow precise automatic contouring of mouse abdominal, thorax, and skeletal structures in under a second.

*Methods.* A DL U-net-like architecture was trained to contour mice organs (GTV, left and right kidney, spinal cord, stomach, liver, bowel, heart, left and right lung, and bones) in small animal radiation research platform (SARRP) CBCT scans. 41 mice were contoured by a human expert, using semi-automatic segmentation methods, after injection of iodine contrast agent, establishing a ground truth for the DL model. The model was trained on a dataset of 26 mice, while 2 mice were used for validation, tuning the model during training, and 15 mice used for performance evaluation testing. The model consists of a pre-processor, and a post-processor for volumetric reconstruction of the DL-predicted probability maps. Model performance was evaluated using both qualitative and distance metrics, including the dice similarity score, precision score, Hausdorff Distance (HD), and mean surface distance (MSD).

*Results.* Performance of the DL-based iodine contrast-enhanced model provided high quality predicted contours in under a second, with the median for all organs being reported: dice > 91%, precision > 95%, $HD_{50}$ < 1.0 mm, and MSD < 1.41 mm.

*Conclusion.* The proposed combination of a DL-based and iodine contrast-enhanced model proved as a viable method to vastly improve efficiency of small animal CBCT image-guided RT preclinical trials.

1. Introduction

The successes of modern radiotherapy (RT) have been mainly technology driven; however, it is expected that future advances will come from using these technologies to pursue biology driven developments (1). Animal models are the backbone of many areas of biomedical and preclinical research and the mouse is the most commonly used model organism for studying human disease (2–5). Understanding and characterizing mouse models in detail is considered key to improving the reproducibility of preclinical results in human subjects (6). For cancer based radiobiology, research guiding clinical treatment often requires preliminary studies on the cellular or small animal scale (7,8). However, to best model the actual radiobiological, radioimmunological, and toxicity characteristics of human therapies, preclinical techniques analogous to modern clinical techniques are needed. Here standardization of preclinical radiation studies using clinically analogous methodologies has recently come to the forefront as a high priority task in moving RT forward (1). As methods in dose calculation, optimization, and image-guided RT (IGRT) continue to be refined, allowing better targeting of radiation to the target, and sparing of nearby healthy tissues, the next step is to leverage these current precise technologies to pursue biology-driven enhancement.

Modern preclinical irradiators have greatly evolved away from the use of radioactive isotopes such as cesium and fixed beam directions to systems that are self-contained in shielded cabinets and use non-radioactive kilo-Voltage (kV) x-ray sources. These systems closely mimic their clinical linear accelerator (linac) counterparts in that the therapeutic radiation source is mounted on a gantry that can be rotated 360° degrees around the mouse to deliver radiation at arbitrary angles. This allows for preclinical intensity-modulated radiation therapy (IMRT) as shown by several recent studies (9–11). Also, like modern linacs, newer irradiators have onboard imaging (OBI) and electronic portal imaging devices (EPID) capabilities. The OBI can capture 2D radiographs as well as tomographic 3D-CBCT images while the mouse is on the treatment table. Such images are used for treatment planning purposes and to verify proper mouse setup with respect to the beam before delivery. EPID images are useful for beam-eye views of the treatment field and can be used for quality assurance (QA) purposes (12).

One of the most time consuming and labor-intensive tasks in preclinical three-dimensional conformal radiotherapy (3D-CRT) is manual organ segmentation. Depending on the site, simple single organ contouring (brain, heart, etc.) can take approximately 20 minutes, whereas, for highly complex cases involving many organs or skeleton contouring for total bone marrow irradiation can take many hours (13,14). Such times are not suitable for large scale murine radiation biology studies which require many mouse deliveries per day to generate statistical significance of results. In addition, manual delineation of organs on each slice of a volumetric CT scan requires not only great attention to detail, but a high-level of mouse anatomy expertise. Furthermore, there is subjectivity involved in manual segmentation due to individual bias.

To address the issue of manual 3D organ segmentation times recent studies have investigated the use of automated segmentation approaches using machine learning. For instance Lappas et al. used a U-net architecture with three levels of encoding to extract features and three levels of decoding to automatically predict organ segmentation (15). This 3D U-net architecture DL model successfully predicted high-quality contours for the thorax and head area in mice µCBCT scans (which has a higher image resolution than the conventional CBCT small animal irradiator). Many researchers have found success using 3D U-net DL architecture for small animal automatic tissue segmentation such as van der Heyden et al., who trained a model to determine skeletal muscle mass in µCBCT scans (16). Moreover, Schoppe et al. used a modified U-net architecture DL model to automatically segment more ambiguously placed and shape organs, such as the pancreas, in µCBCT scans, with and without contrast-enhancing agent (17). In these examples, however, a higher resolution µCBCT irradiator is used, as opposed to the more commonly available conventional CBCT irradiator, making it easier to segment due to higher image contrast. In this paper, we demonstrate the capability of using a machine learning model with conventional CBCT irradiators by combining DL U-net-like architecture with iodine-based contrast enhancement to successfully predict organ segmentation efficiently and accurately in larger more ambiguously shaped organs.

Although AI has shown great potential for automated organ segmentation in µ-CT images, the lower quality of preclinical irradiator CBCT had limited AI to only areas that are high in contrast, and largely considered unambiguous for segmentation specialists, such as the heart, lungs, and skeleton (15). This greatly limits the use of AI organ segmentation for most preclinical RT studies. To address this issue, and to further expand the utility of AI contouring in preclinical RT, we present a method of using AI-enabled mouse organ segmentation, combined with iodine-based contrast enhancement, to efficiently and accurately contour unambiguous organs in the thorax and bones, as well as, mostly ambiguous, abdominal mouse organs. To the best of our knowledge, this is the first method to perform high quality DL-generated abdominal organ contours in lower quality CBCT scans and has the potential to vastly improve efficiency of small animal image-guided RT preclinical trials. Moreover, this study addresses the feasibility of DL-enabled heterogonous structures, such as the GTV.

## 2. Methods and Materials
### 2.1. Dataset and Image Conditioning

The dataset consists of 41 6-week-old immunosuppressed CD-1 nude mice partly taken from a previously completed pancreatic cancer study (18). In this study, treatment planning and irradiation of the mice were performed when the gross tumor volume (GTV) reached 90 mm$^3$, which took about eight weeks post injection of Panc-1 (CRL-1469) and three weeks for MiaPaCa-2 (CRL-1420). These two human pancreatic carcinoma cell lines were obtained from the American Type Culture Collection. Mice were injected with $2.0 \times 10^6$ cells for Panc-1 or $1.5 \times 10^6$ cells for MiaPaCa-2 into the parenchyma of the pancreas. Anesthesia was used during the entirety of the imaging process to immobilize the mice. Each mouse inhaled isoflurane anesthesia at a concentration of 1.5% with a 6% volume of oxygen as a carrier of the gas. Imaging was done using cone-beam computed tomography (CBCT) in a SARRP device, with the operating x-ray source at a voltage of 60 kV and current of 0.8 mA. In addition, each mouse was injected with 5 mL per 1 kg mice body weight of iodine-containing contrast agent Imeron intravenously (iv) through the lateral tail vessel to improve soft tissue contrast and optimize the definition of the tumor volume, as well as the organs at risk (OARs). CBCT imaging was taken immediately after Imeron iv injection using 1440 projections and a voxel size of 0.115 x 0.115 x 0.115 mm$^3$.

Treatment planning, including manual segmentation of mice organs, which were mostly done using the preclinical treatment planning software 'MuriPlan'. The GTV and six OARs, including both kidneys, small bowel, stomach, spinal cord, and liver, were manually delineated slice by slice of the CBCT image by a segmentation specialist. The GTV was defined as the macroscopically visible tumor tissue within the pancreas, with a safety margin of 2 mm surrounding the GTV being considered. Contouring of the heart, lungs, and bones was done semi-automatically, using a previously made model to predict contours, then carefully going slice-by-slice to fix any errors. Each mouse has been professionally contoured slice by slice to delineate the GTV (the tumor present in the pancreas), both kidneys, small bowel, stomach, spinal cord, and liver (18). As previously mentioned, as a means of optimizing detection of organs and orthotopic tumor in the pancreas, iv injection of Imeron was used immediately before imaging to increase soft tissue contrast. This allows the pancreatic tumor to appear hypodense in the CBCT image. In addition, the intestinal loops are partly air-filled, the kidneys in the retroperitoneum contain strong contrast in the presence of Imeron, and the spine and all other skeleton structures appear hyperdense. Moreover, the correct identification of the orthotopic tumor and other OARs in CBCT images was validated and confirmed via gamma-H2AX staining (18).

As a means of standardizing the Nifiti CT images before they are read by the DL model, are put through various preprocessing measures. It is important that both the scan and annotated 3D Nifti volumes are of the same size and that in the 'ground truth' file, which is the manually segmented volume, each voxel is encoded with an integer that represents a specific organ in the segmentation class. The script first checks for completeness of the available scans and verifies them for consistency. Then, to encode the organs a set is used to pair a string to an integer. Using the pickle module, which implements a binary protocol for serialization of a Python object structure, 'pickling' occurs for each 3D Nifiti volume. 'Pickling' is a process that converts a Python object hierarchy into a byte stream (19). This means that the given variable (a string describing the organ) is saved into a file which the DL model will use to identify each organ by deserialization of the pickle file ('pickdump'). Finally, each 3D volume is turned into coronal slices to allow for faster processing. These 2D Nifti files get sliced into hundreds of tiff files and the signal intensity is normalized by subtracting the mean and dividing by the standard deviation. The images then get resampled to a desired resolution of 240 µm/vx, even though the DL AIMOS U-Net-like architecture works over a broad range of 120 – 1120 µm/vx (17).

### 2.2. Devices and Python Libraries

The algorithm, developed in python, was implemented using these open-source libraries: PyTorch, SciPy, NiBabel, Nrrd, Numpy, and Matplotlib. The deep learning framework pipeline uses PyTorch. Additionally, a NVIDIA Titan Xp GPU was used, but not necessary as the program can be run using a CPU, at the cost of time. The pipeline, proposed by (17), consists of three parts, a preprocessor, deep learning backbone, and a postprocessor.

### 2.3. Deep Learning Model

The DL backbone follows a U-Net-like architecture, consisting of an encoding and decoding path connected with skip connections (17). Skip connections are used for images of varying sizes. The varying convolutional sizes help to extract features at various resolutions and without skip connections, upsampling back to the original resolution is very difficult. Moreover, skip connections resolve the vanishing gradient problem by providing an uninterrupted gradient flow from the first to last layers (20). The number of encoding and decoding levels is varied, but always follows the same structure. Encoding levels consist of two convolutions (kernel size: 3, padding: 1, stride: 1), batch normalization, a rectifying linear activation function (ReLu), and a max-pooling operation (kernel size: 2, stride: 2). Additionally, each encoding level has twice the number of featured channels compared to its previous level, starting at 32 feature channels. The decoding units consist of three convolutions (kernel size: 3, padding: 1, stride: 1) and receive concatenation of upsampled input from the previous level, including input from the skip connection form the corresponding encoding level (consisting of the same number of feature channels). The final convolution maps the 32 feature channels to the number of organs to be predicted. The last pass is through a sigmoid function which creates a volumetric probability map for each organ.

The DL model is trained with a soft-Dice loss function and uses the Adam optimizer (17). A soft-Dice loss function is used to directly predict probabilities of organ position, as opposed to converting them into a binary mask. The soft-Dice loss function is a differential reformation of the commonly used Dice Similarity Coefficient (DSC) evaluation metric. If $A$ represents pixels from the human segmentation ground truth file, and $B$ consists of pixels from the DL automatically segmented file, the DSC can be calculated as two times the intersection of A and B divided by the sum of the two sets: $DSC = \frac{2|A \cap B|}{|A| + |B|}$ (21). In realizing that $2|A \cap B| = \sum_i a_i * b_i = <A,B>$ and $|A| = \sum_i a_i * a = <A,A>$, a differential soft-Dice loss function can be determined as:

$$L_{Dice} = 1 - \frac{2<A,B>}{<A,A> + <B,B>}$$

The Adam optimizer was proposed by Kingma and Ba (22), as a method for computing individual adaptive learning rates for different parameters from estimates of first and second moments of the gradients. Adam is an efficient stochastic optimization that uses first-order gradients with little memory requirement. Furthermore, a nested k-fold cross-validation procedure is used to split the dataset into a training set (for model weight optimization), a validation set (for hyper-parameter optimization), and a test set (for evaluation) (22). The DL model is trained for 30 epochs, with an initial learning rate of $10^{-3}$, on the training set, with the learning rate being gradually reduced as validation performance is unchanged over five epochs. The post processor utilizes something known as ensemble dash voting, which means the neural network uses information from many independently trained networks and merges them to make the most accurate prediction possible (17). Finally, the volumetric probability map from the DL backbone is used for volumetric reconstruction to create a predicted segmentation file in the postprocessor.

### 2.4. Evaluation Metrics

Both quality and distance metrics are presented for performance evaluation. Quality metrics, such as the Dice Similarity Score, or just dice score, is an evaluation of the qualitative performance of the algorithm with respect to the human segmentation for each organ. Quality metrics use voxel classification of the DL-predicted contouring of a SARRP CT scan for each organ, denoting each voxel as either 'True Positive' (TP), 'False Positive' (FP), or 'False Negative' (FN). Voxels are classified as TP if it was correctly predicted compared to the ground truth file, FP if the voxel was incorrectly predicted as part of an organ, and FN if the voxel was predicted to be a non-organ but is a part of an organ in the ground truth file. With this classification of the pixels, additional metrics can be utilized to assess accuracy of automatically segmented files against human segmentation, such as those presented in the Table 1. The quality metrics are based on two sets of pixels, A and B, which consists of pixels from the DL-predicted segmentation and the human ground truth segmentation, respectively. The dice index gives the percent overlap, or reproducibility–how well the segmentation and ground truth files match–for each organ. The Jaccard index is the intersection of A and B over the union of A and B. The precision gives the rate at which pixels are correctly classified as part of the correct organ. FP rate and FN rate give the rate at which false positives classifications are found and false negative predictions are found, respectively (23).

Spatial distance metrics are used to quantitatively evaluate the accuracy of the predicted contours, taking into consideration the spatial position of voxels. The directed Hausdorff (h(A,B)), as shown in the Table 1, where |a-b| is the difference in the Euclidean distance between pixels in their respective pixel sets from predicted contouring of an organ and ground truth contours. The Hausdorff Distance (HD) is the distance between crisp volumes (two finite pixel sets, A and B). Since noise and outlier pixels are very common in medical segmentation, it is not recommended to use the maximum HD. Instead, the quantile method proposed by (24) is one way to handle outliers, where HD is defined as some percentile of the maximum distance (23). For this reason, the 50th and 95th percentile HDs are presented. The mean surface distance gives the mean distances between predicted and human segmentations for each organ. The median and standard deviation of these distances are also reported.

| Quality Metrics Formula (23) | Distance Metrics Formula (25) |
|---|---|
| $Dice\ Score = \dfrac{2TP}{2TP\ +\ FP\ +\ FN\ +\ \epsilon} \times 100\%$ | $h(A, B) = max_{a \in A} min_{b \in B} |A - B|$ <br> $HD(A, B) = max(h(A, B), h(B, A))$ |

| | |
|---|---|
| $Jaccard\ Index = \dfrac{TP}{TP + FP + FN + \epsilon}$ | $HD_p(A, B) = percentile[max(h(A,B), h(B,A))]$ |
| $Precision = \dfrac{TP}{TP + FP + \epsilon}$ | $Mean\ SD = mean(h(A,B), h(B,A))$ |
| $FP\ rate = \dfrac{FP}{FP + TP + \epsilon}$ | $Median\ SD = median(h(A,B), h(B,A))$ |
| $FN\ rate = \dfrac{FN}{FN + TP + \epsilon}$ | $std\ SD = std(h(A,B), h(B,A))$ |
| $\epsilon = 0.0001$ for numerical stability | SD = Surface Distance, HD = Hausdorff Distance |

*Table 1. Quality metrics used for performance evaluation of the DL model (left) and distance metrics used for evaluation of DL model (right).*

### 2.5. Visualization Techniques

As a means of representing the DL predicted segmented SARRP data, both qualitatively and quantitatively, three separate visualization techniques were developed. The first qualitative visualization technique employed uses the napari python library. Napari is a self-described fast interactive viewer for multi-dimensional image representation. The predicted segmentation python array is loaded into the napari interface, allowing the user to easily evaluate each individual slice of the predicted segmentation (26). The next qualitative visualization technique utilizes interactive intensity plots. The three plots were developed, showing a side-by-side comparison of all the slices of the deep learning-predicted segmentation, the ground truth segmentation file, and the difference between the two. The final visualization technique utilizes the Dice score to provide quantitative information on the AI predicted segmentation. This plot shows each individual organ with a highlighted section showing the predicted segmentation of the organ. Above each individual organ plot is the Dice score, showing how well the predicted segmentation matches the ground truth file.

### 3. Results

The DL model was trained on a set of 26 mice. Of the 26 mice used, one mouse was used for testing and one mouse was used for validation, during training of the DL model. The model was trained on 30 epochs and tested on a test set of 15 mice. Figure 1 shows representatively a side-by-side comparison of 3D projections of human segmented ground truth file (left) and AI automatically segmented file (right). As shown, the two look remarkably similar, which has extremely positive implications for the efficacy of the DL model.

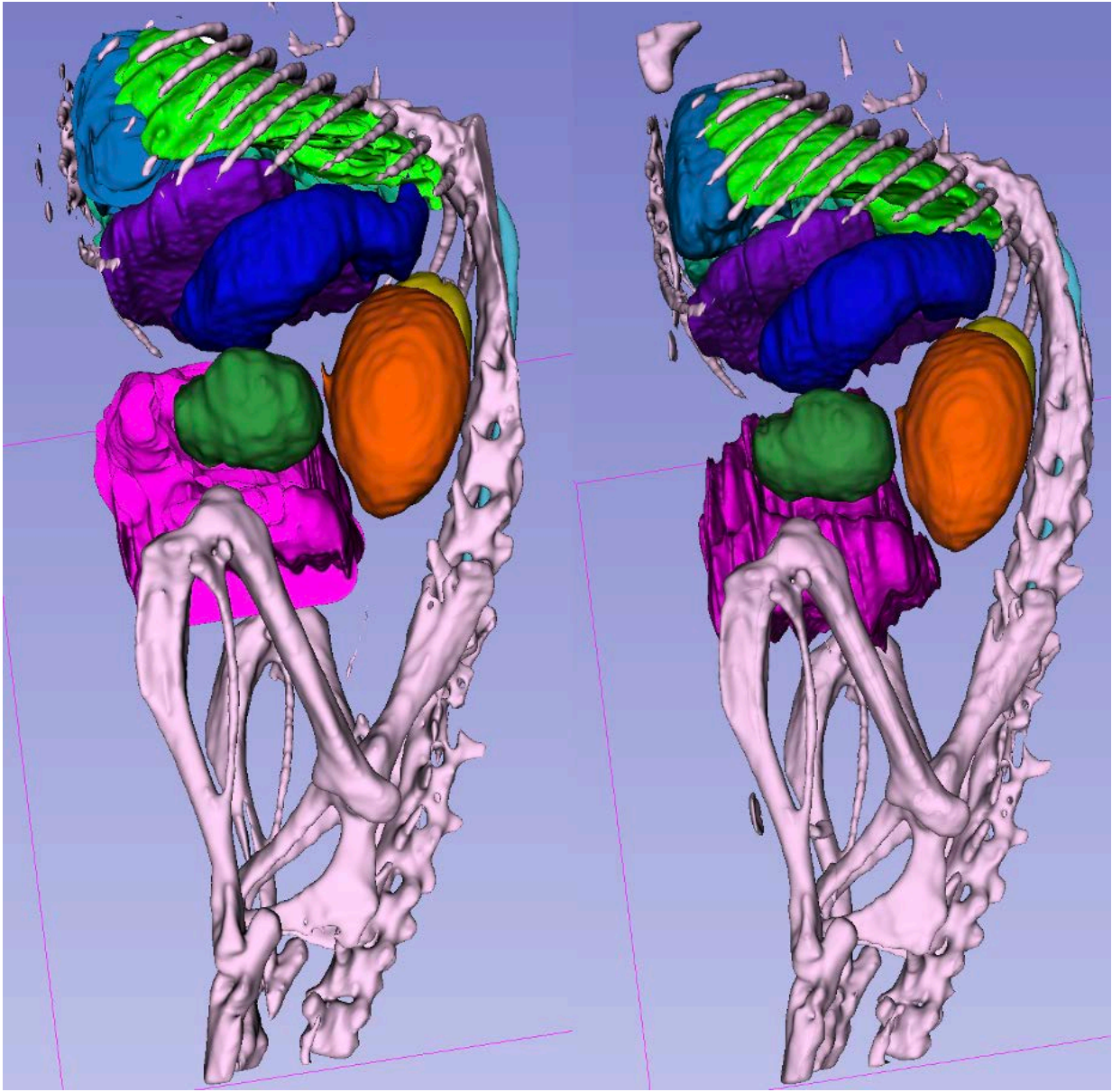

*Figure 1: Mouse data: mouse_17. Human ground truth 3D projection of the segmentation file (left) juxtaposed to the AI-contoured file (right). The GTV is dark green, left kidney is orange, right kidney is gold, spinal cord is aqua blue, bowel, stomach is blue, liver is purple, bowels are magenta, heart is light blue, left lung is bright green, right lung is chartreuse, and the bones are pink.*

As suggested by Schoppe et al. (17), it was found that training over 30 epochs produced the best results. Additionally, the ideal training dataset size was found to be 26 mice. For this reason, a 26-mouse test dataset was used, with one of the 26 mice being used for testing during training, and another one used for validation for the Adam optimizer during training. This left the remaining 15 mice to be used for evaluation of the model.

For the best contours, the DL generated segmentations for each organ almost perfectly match the human ground truth contours, while in the worst case there are only few regions of disagreement. Figure 2 shows example slices of the coronal and sagittal views for one mouse from the test dataset, with the ground truth contours of the organs outlined in bolded yellow. The DL generated contours are shown highlighting each predicted organ. Areas of disagreement come in the form of either a pixel characterized as not part of a specific organ, when the ground truth file had that pixel as a part of that organ, or a pixel is characterized as part of a specific organ by the DL model, while that pixel is not part of the organ in the ground truth file. The former instance of disagreement is depicted as the space between the yellow outline and the highlighted organ, whereas the latter instance of disagreement is depicted as the highlighted organ outside of the closed surface ground truth yellow outline.

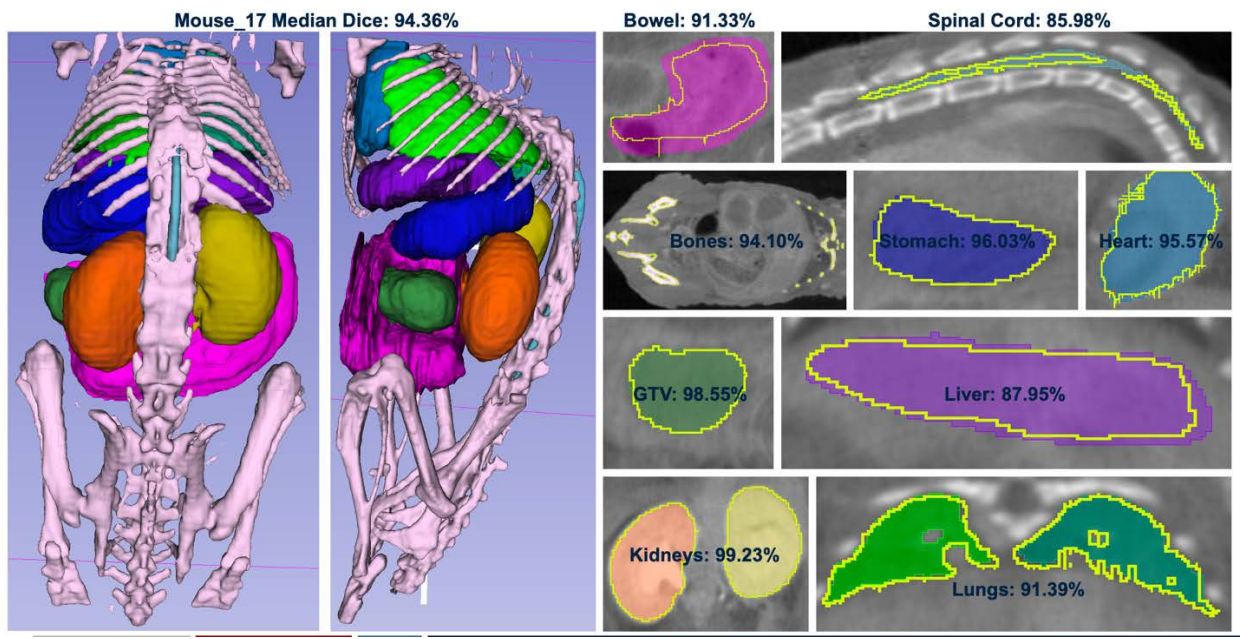

*Figure 2: 2D slide depictions in the coronal, and sagittal views for one mouse in the 15-mouse testing set. The ground truth human segmentations are shown as a bold yellow outline. AI automatic segmentations are shown in the highlighted colors. The GTV is dark green, left kidney is orange, right kidney is gold, spinal cord is aqua blue, bowel, stomach is blue, liver is purple, bowels are magenta, heart is light blue, left lung is bright green, right lung is chartreuse, and the bones are pink.*

Figure 3 displays the dice score boxplots results for each organ, as well as the median of all organs, for all 15 mice used for evaluation. Median dice scores for each organ: GTV = 91.86%, left kidney = 99.06%, right kidney = 99.13%, spinal cord = 87.07%, stomach = 95.24%, liver = 87.95%, bowel = 89.65%, heart = 94.11%, left lung = 90.10%, right lung = 91.21%, bones = 87.43%, and average for all organs = 91.21%. The median dice score is higher than 85%, meaning overall performance of the DL automatic segmentation model is very high (15). The kidneys show consistent accuracy of over 99%, whereas the stomach, liver, and bowel, are slightly less accurate, but still consistently over 87%. Segmented structures in the thorax region, as expected, performed very high, consistently over 90%. Nevertheless, where the model lacks in accuracy is longitudinally, in the spinal cord and skeleton, for instance, where there is high variance. The accuracy of the spinal cord contours is still very good, averaging to about 87% dice score. The median dice score for the bones, however, show even more variance, ranging from 60% dice score, all the way up to 95%, with a median of about 87%. Thus, the worst contours were only as bad as a 87% dice score, which is highly accurate. Although the GTV is heterogenous, the DL model performed very well predicting these structures, ranging from dice scores of about 77% to as high as over 98%.

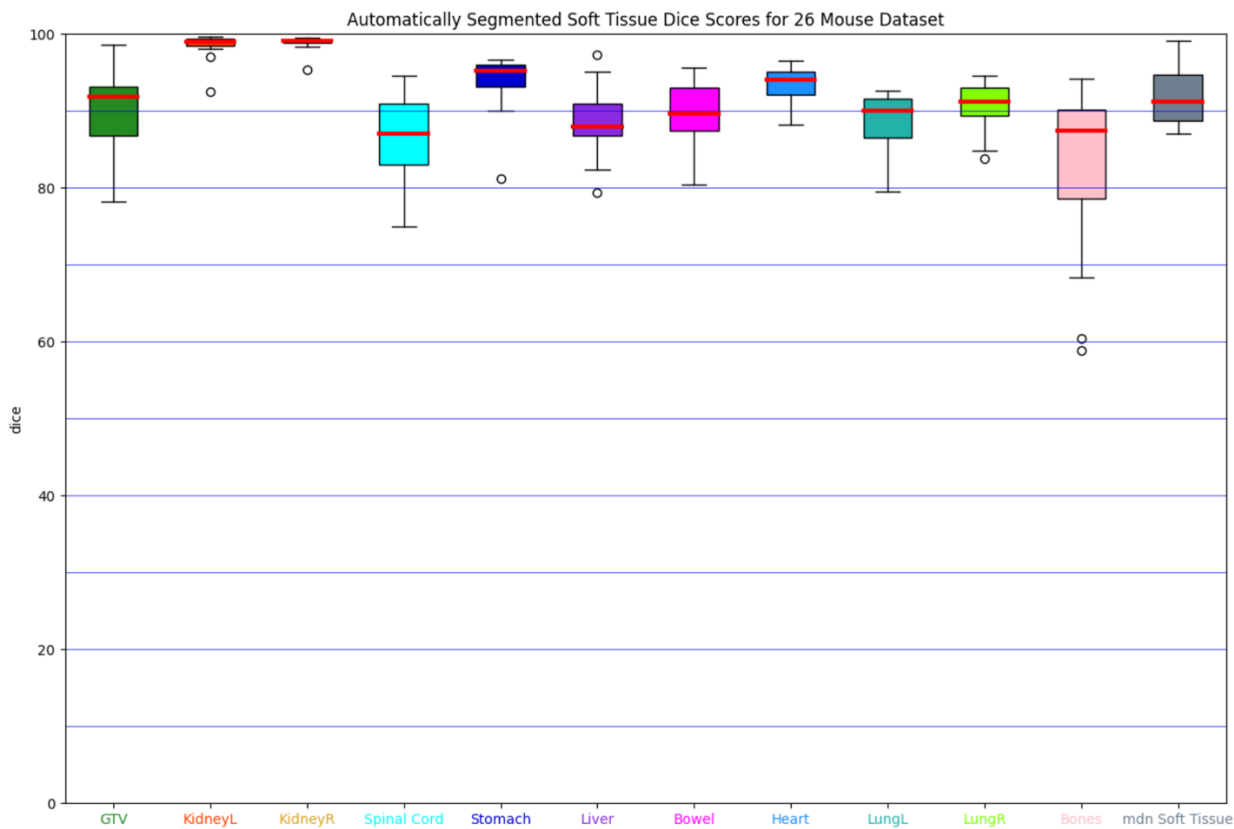

Figure 3: Dice scores of DL generated contours compared to manual human ground truth contours. Boxplots show the median (solid red line), interquartile range (top and bottom of the boxes), 1.5 times the interquartile range (whiskers), and outliers (circles). The GTV is dark green, left kidney is orange, right kidney is gold, spinal cord is aqua blue, bowel, stomach is blue, liver is purple, bowels are magenta, heart is light blue, left lung is bright green, right lung is chartreuse, and the bones are pink.

For all eleven annotated structures–GTV, left kidney, right kidney, spinal cord, stomach, liver, bowels, heart, left lung, right lung, and bones–each evaluation metric is presented in Table 2. Similar trends in the results can be observed in the Jaccard and precision scores. According to the Jaccard scores, the least accurate segmentations came in the spinal cord, liver, and bones, but still consistently over 77%. The model produced extremely good precision scores, with the least precise predicted volume segmentation came in the heterogenous GTV, with a precision score of 88.51%. The false negative rate (FNR) or miss rate of the DL-enabled mouse automatic segmentation model for each organ is shown in the fifth column of Table 2. Overall, there is a median miss rate of about 8%, which shows high accuracy of the DL automatic contours. The 50th percentile Hausdorff Distance (HD) averaged for all organs is about 1.00 mm, while the 95th percentile HD averaged for all organs is about 5 mm. The average mean surface distance for all organs is about 1.41 mm. The average median surface distance for all organs is about 1.00 mm. The standard deviation for all organs averaged to about 1.82 mm. Overall, distance measurements show strong agreement with quality metrics, showing high accuracy of DL automatic contours. According to the surface distance measurements, consistently, the least accurately contoured organ is the bowel. This is because the bowel is the largest area to contour, as the bowels are spread along the whole abdomen. The bones, however, have the largest miss rate at 20%, as shown in Table 2.

|  | Median of Metric | | | | | | | | |
|---|---|---|---|---|---|---|---|---|---|
| Organ | Dice | Jaccard | Precision | FNR | $HD_{50}$ | $HD_{95}$ | Mean SD | Median SD | Std. SD |
| GTV | 91.86 | 84.94 | 88.51 | 0.07 | 1.00 | 3.61 | 1.41 | 1.0 | 1.16 |
| Left Kidney | 99.06 | 98.15 | 98.82 | 0.00 | 0.00 | 2.0 | 0.26 | 0.0 | 0.65 |
| Right Kidney | 99.13 | 98.28 | 99.48 | 0.01 | 0.00 | 2.0 | 0.25 | 0.0 | 0.55 |
| Spinal Cord | 87.07 | 77.11 | 90.0 | 0.08 | 0.00 | 2.0 | 0.53 | 0.0 | 0.88 |
| Stomach | 95.24 | 90.91 | 95.21 | 0.05 | 1.0 | 3.16 | 1.17 | 1.0 | 1.17 |
| Liver | 87.95 | 78.49 | 90.01 | 0.08 | 2.00 | 8.0 | 3.07 | 2.0 | 2.30 |
| Bowel | 89.65 | 81.24 | 91.23 | 0.09 | 3.16 | 14.87 | 5.70 | 3.16 | 4.89 |
| Heart | 94.11 | 88.87 | 97.93 | 0.10 | 1.0 | 5.95 | 1.82 | 1.0 | 1.87 |
| Left Lung | 90.10 | 81.99 | 98.97 | 0.17 | 1.0 | 6.32 | 1.52 | 1.0 | 1.99 |
| Right Lung | 91.21 | 83.84 | 96.41 | 0.09 | 1.0 | 5.00 | 1.36 | 1.0 | 1.82 |
| Bones | 87.41 | 77.63 | 92.81 | 0.20 | 1.0 | 6.00 | 1.80 | 1.0 | 2.25 |
| All | 91.21 | 83.84 | 95.21 | 0.08 | 1.0 | 5.00 | 1.41 | 1.0 | 1.82 |

*Table 2: Segmentation accuracy evaluated using the dice, Jaccard score, precision score, FNR, 50th percentile and 95th percentile HDs, mean, median, and standard deviation of surface distances. The values displayed are the medians for each organ over the while 15-mice test dataset.*

Figure 3 displays a comparison of CBCT scans of a native, non-contrast-enhanced mouse, and a contrast-enhanced mouse. As shown in the axial (top left), coronal (bottom), and sagittal (top right) slices, there is a clear advantage in using iodine-based contrast-enhancements. When inputting a native CBCT scan into the DL-based automatic segmentation model, performance is very poor, and limits quality contours to only the thorax region, as well as the bone structure. For this reason, the use of contrast-enhancement immediately before CBCT imaging is imperative to building an accurate abdominal organ automatic segmentation model, with this U-net-like DL architecture.

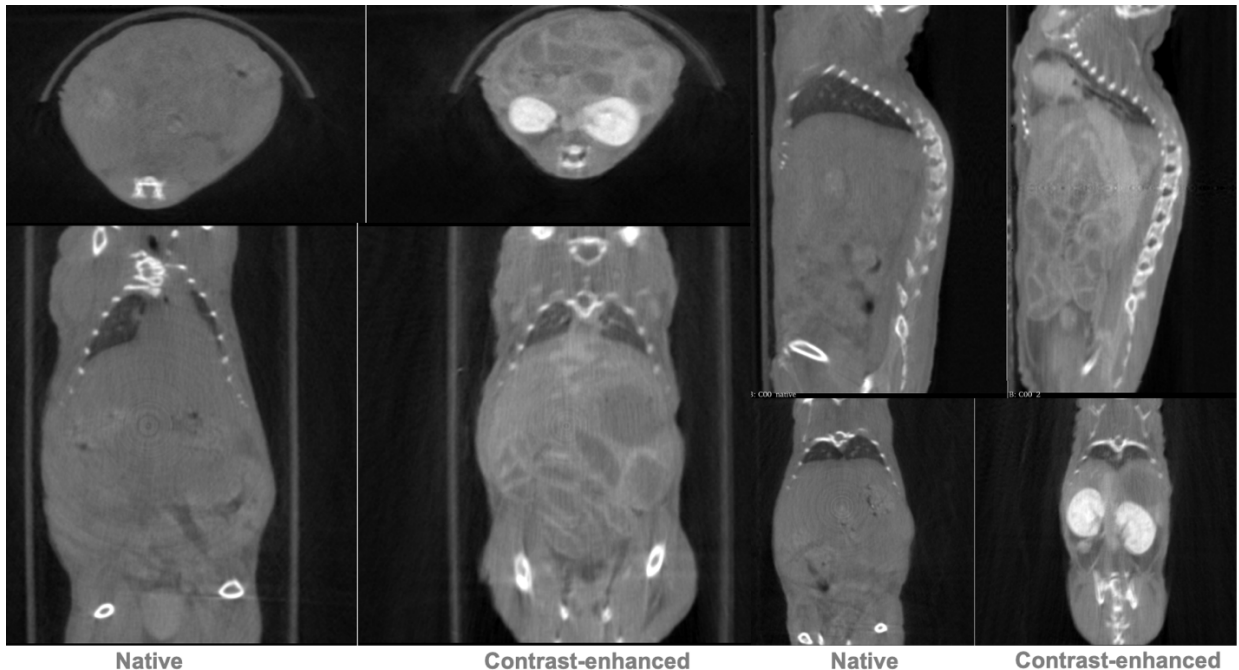

*Figure 3: Comparison of non-contrast-enhanced, native, CBCT scan versus iodine-based contrast-enhanced CBCT scan. Contrast-enhancement causes the GTV in the pancreas to appear hypodense, intestinal loops are partly air-filled, kidneys in the retroperitoneum, and skeletal structures appear hyperdense.*

## 4. Discussion

Despite image resolution limitations of conventional, commonly used, small animal CBCT irradiators, the combination of iodine-based contrast-enhancement and U-net-like DL architecture provided high-quality, efficient, contours of abdominal organs, without affecting accuracy of thorax and bone structures. Results presented in this paper have positive implications for future preclinical 3D conformal RT trials. Not only does DL-enabled automatic organ segmentation vastly increase efficiency of these periclinal murine studies, but it also eliminates human bias and fatigue. Moreover, the addition of iodine-based contrast-enhancement decreases ambiguity in abdominal organ areas, increasing the quality and precision of the DL-enabled contours.

Small animal irradiators have come a long way in recent years to best match the clinical irradiators as closely as possible. For instance, the two most common commercially available small animal irradiators, SARRP system (Xstrahl Ltd, UK) and X-RAD 225Cx (Precision X-ray Inc., USA), are equipped with advanced OBI, EPID, and treatment planning systems (TPS). However, critical issues still remain, such as the largely understudied aspect of intra-irradiation motion and photon scattering for narrow beams of kV energies (27). In particular, the latter may cause issues with CBCT imaging quality and dose calculations. Conventional CBCT images are accomplished using a fixed X-ray tube/imager, while micro-CTs use a rotating gantry system. Mechanical hardware accuracy has been reported to be below 100 µm, while typical image resolution is in the 100-200 µm range, and even lower for µ-CTs (27). The large time requirement for tissue segmentation is mainly dependent on anatomical complexity (areas considered ambiguous for segmentation specialists). However, the use of contrast enhancing agent can vastly decrease anatomical complexity, revealing relevant features in tissue that were previously non-visible, as shown in Figure 3 (28). By iv injection of iodine-based contrast-enhancement, Imeron, immediately before CBCT imaging, the GTV appeared hypodense, intestinal loops partly air-filled, kidneys in the retroperitoneum, as well as the spine and other skeletal structures appearing hyperdense (18). These contrast enhancements made human ground truth segmentation possible for the radiation biologist, allowing for a novel DL-enabled contours.

The results from this study are consistent with previously established findings in this research area, showing high precision and usefulness of DL-based automated contouring over conventional approaches (15). Moreover, this study demonstrates the capability of DL-based abdominal, thorax, and skeletal organ structure segmentation in conjunction with iodine-based contrast-enhancement in conventional CBCT scans. Manual organ segmentation takes upwards of 10 minutes per organ, making the total segmentation time in this study over 75 hours (15). Conversely, the DL model can accurately segment all eleven mouse organs and skeletal structures in under a second, vastly reducing contouring time and increase efficiency of the image-guided precision RT workflow. The U-net-like DL architecture proved as a method to accurately, automatically contour murine abdominal and thorax organs, as well as skeletal and heterogenous tumor structures, including the GTV, right and left kidney's, spinal cord, stomach, liver, bowel, heart, left and right lung's, and bones in under a second. Upon training on a dataset of manual expert and semi-autonomous segmentation, performance of the test mice for most of the organs showed high quality automatic contours, consistently way over a dice score of 85% and mean HD of about 1.0 mm. In addition, the DL architecture in conjunction with contrast-enhancement is capable of predicting the GTV, with extremely high accuracy.

There are many sources for uncertainty in this DL-enabled automatic contouring architecture that influences the quality of the generated segmentations (15). Most notably, there is uncertainty that can arise in the input. Manual, human contours may not represent a true ground truth segmentation due to subjective interpretation and bias, which must be taken into account (17). This subjective interpretation effect is quantified using inter-observer variability (IOV) studies, which potentially causes regions of disagreement. Additionally, the observer, or scientist performing the manual segmentation of organs may suffer from fatigue and time pressure, causing them to miss contours over time, causing incompleteness, causing disagreement as well. Limitations of using DL for automated delineations of small animal organs mainly stem from CBCT artifacts causing the model to predict pixels that are a part of the CT bench as part of an organ. This effect, however, can be quickly and easily resolved in most treatment planning software by simply deleting the contours that are clearly out of the image of the small animal.

The model lacked in accuracy mainly in areas often considered ambiguous for manual segmentation specialists. The metrics presented are extremely sensitive and consider all differences between the DL automatic contours and human ground truth segmentation contours. Thus, relatively poorer scores can be observed longitudinally in the spinal cord or skeletal structures, as found in previous studies. Additionally, inaccuracies are most common in the larger structures, such as the bowel, which is hard for the segmentation specialist as well.

## 5. Conclusion

In this work, a DL model was successfully trained to accurately predict contours in the abdominal and thorax region in contrast-enhanced mouse SARRP CBCT scans in under a second. This program is dedicated to preclinical work, and in particular, a proposed method to draw strong parallels between preclinical and clinical highly conformal RT studies. In terms of DSC, the median for all organs reported scores over 91%, with over 95% precision, as well as one-millimeter mean HD. Future work includes further insight into the uncertainty of the model by acquiring organ contours from another annotator (segmentation specialist) to quantify inter-observer variability (IOV).